\documentclass[letterpaper, 10 pt, conference]{ieeeconf}  

\IEEEoverridecommandlockouts                              
\usepackage{multirow}
\usepackage{booktabs}  
\usepackage{array}         

\usepackage{graphicx}        

\usepackage{hyperref} 
\usepackage{hyperref}
\usepackage{url}
\usepackage{algorithm}
\usepackage{algorithmic}
\usepackage{multirow}
\usepackage{multicol}
\usepackage{booktabs}
\usepackage{graphicx} 
\usepackage{colortbl}
\usepackage{makecell}  
\definecolor{mygray}{gray}{.9}
\definecolor{myblue}{rgb}{0.9, 0.9, 1}
\let\labelindent\relax
\usepackage{enumitem}
\usepackage{wrapfig}                                                         
\usepackage{graphicx}
\usepackage{microtype}
\usepackage{algorithm}
\usepackage{algorithmic}
\usepackage{newfloat}
\usepackage{listings}
\usepackage{booktabs} 
\usepackage{multirow}
\usepackage{pifont}  
\usepackage[style=numeric, sorting=none]{biblatex} 
\usepackage{amsmath}
\usepackage{amssymb}
\usepackage{xcolor} 

\usepackage{inconsolata}

\usepackage{graphicx}
\usepackage{caption}
\title{\LARGE \bf
AdaToken-3D: Dynamic Spatial Gating for Efficient 3D Large Multimodal-Models Reasoning
}

\author{
 Kai Zhang $^{1}$, Xingyu Chen$^{3}$ and Xiaofeng Zhang$^{2}$ 
}

\begin{document}

\maketitle
\thispagestyle{empty}
\pagestyle{empty}

\begin{abstract}

Large Multimodal Models (LMMs) have become a 
pivotal research focus in deep learning, demonstrating
remarkable capabilities in 3D scene understanding. 
However, current 3D LMMs employing thousands of 
spatial tokens for multimodal reasoning suffer from 
critical inefficiencies: excessive computational 
overhead and redundant information flows. Unlike 2D
VLMs processing single images, 3D LMMs exhibit
inherent architectural redundancy due to the heterogeneous
mechanisms between spatial tokens and visual tokens. To
address this challenge, we propose AdaToken-3D, an adaptive
spatial token optimization framework that dynamically prunes
redundant tokens through spatial contribution analysis. Our 
method automatically tailors pruning strategies to different 3D 
LMM architectures by quantifying token-level information flows 
via attention pattern mining. Extensive experiments on LLaVA-3D (a 7B parameter 3D-LMM) demonstrate that AdaToken-3D 
achieves 21\% faster inference speed and 63\% FLOPs reduction 
while maintaining original task accuracy. Beyond efficiency 
gains, this work systematically investigates redundancy patterns 
in multimodal spatial information flows through quantitative 
token interaction analysis. Our findings reveal that over 60\% of spatial tokens contribute minimally ($<$5\%) to the final predictions,
establishing theoretical foundations for efficient 3D multimodal
learning.

\end{abstract}

\section{INTRODUCTION}

 Large Multimodal Models (LMMs) have demonstrated 
remarkable capabilities in 2D visual-language tasks. Recent 
advancements in integrating multimodal 3D representations including RGB-D data \cite{3,4}, point clouds \cite{5,6} have empowered 
LMMs with unprecedented 3D scene understanding capacities. These enhanced models now demonstrate proficiency in
complex 3D reasoning tasks such as visual grounding and
dense captioning within three-dimensional environments. Their emergence represents a paradigm shift that 
fundamentally expands the operational boundaries of large 
language models. This technological leap is widely recognized as a critical pathway toward achieving human-level spatial 
intelligence and advancing the development of embodied AI 
systems. Spatial tokens serve as gateways for 3D scene understanding. Similar to 2D LMMs that employ non-linear mappings from pixels to visual tokens, 3D LMMs typically integrate spatial tokens encoded from point clouds with multimodal features. Individual spatial tokens encapsulate latent patterns across entire 3D environments, enabling 3D scene understanding. However, the iterative computation of spatial tokens in current architectures exhibits critical memory inefficiencies and reasoning redundancy. As demonstrated in Fig\textcolor{red}{.1}, thousands of spatial tokens make negligible contributions to deeper layers, revealing 
information redundancy in 3D LMMs. This redundancy persists in large-scale fused representations combining visual tokens with 3D positional embedding.
\begin{figure*}[htpb]
    \centering
    \includegraphics[width=2\columnwidth]{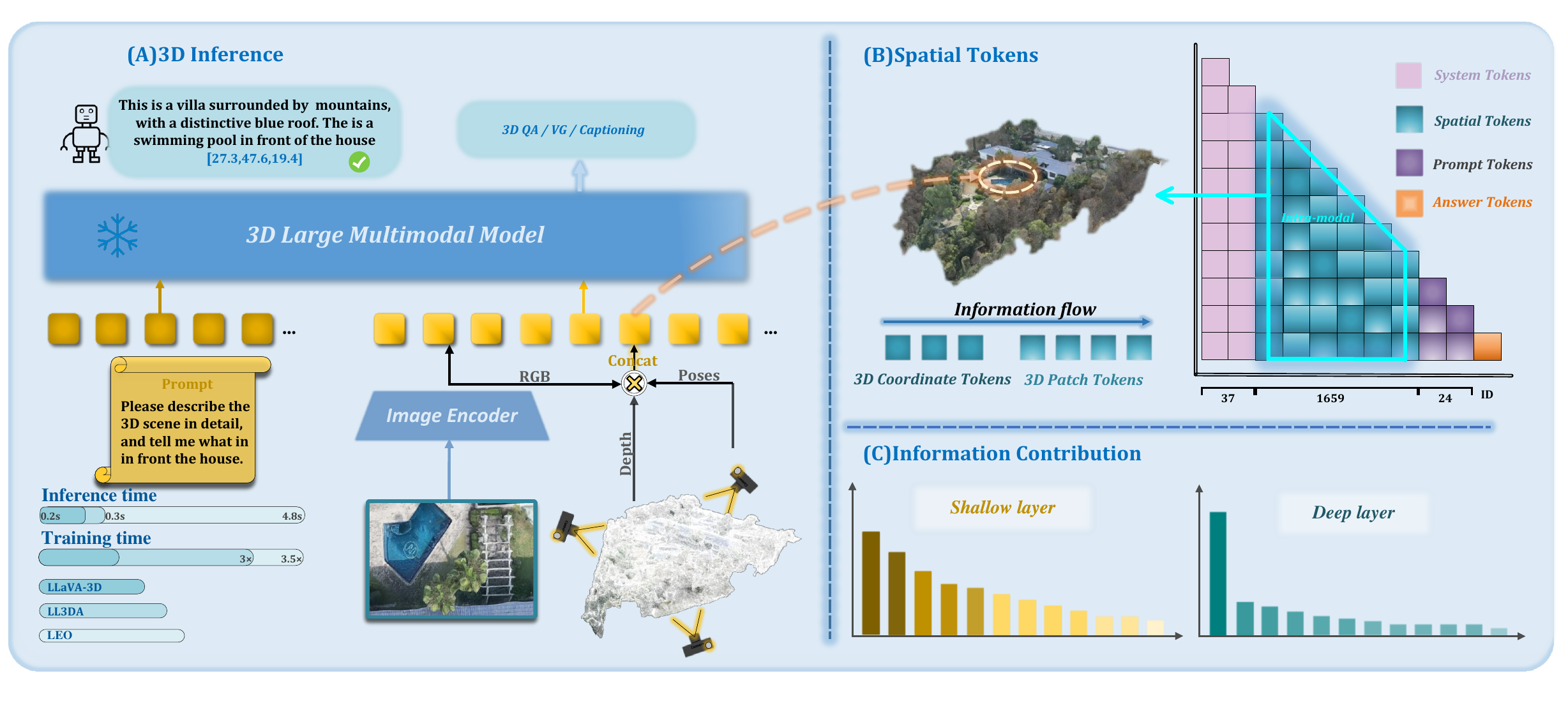}
    \caption{Overview of Spatial tokens in 3D LMMs. Block (A) demonstrates that LLaVA-3D is injected with 3D patches to acquire the 3D scene understanding while maintaining the lightweight architecture as 2D LLMs. LLaVA-3D could achieve state-of-the-art performance across a wide range of 3D benchmarks(3DQA 3DVG). The right Blocks (B) and (C) have pointed out the spatial tokens id index and different types of token processed in LLaVA-3D. Then comparing the information contribution of various tokens in different layers. }
    \label{figure :st_collision_type}
\end{figure*}

Previous work has explored the information flow and 
reasoning mechanism in LVLMs. Information Flow In 
LVLMs\cite{1} creatively proposes a novel analysis method for  
2D LLMs and discovers the information flow converges in 
shallow layers while diverging in deeper layers. FastV\cite{2}
approaches the problem from the angle of computing the 
attention of 2D visual tokens. As multimodal visual 
information is injected into 3D LMMs,  whether it still follows 
the reasoning mechanism of LVLMs needs new experimental  
verification. To this end, we observe the spatial information 
flow in deep layers of  efficient 3D LMMs, such as LLaVA-3D, and introduced Information Contribution to quantify the 
information flow of 3D LMMs. Therefore, our  research is 
mainly aimed at solving two problems:

\textbf{Q1:How to optimize reasoning in the case of redundant spatial tokens and maintain spatial  task performance of 3D LMMs?}

\textbf{Q2:What are the characteristics of spatial information flow in 3D LMMs?}

While recent attention-based methods for accelerating inference have demonstrated significant improvements in the efficiency of Large Vision-Language Models (LVLMs) \cite{7,20,21}, it is important to recognize that, unlike individual image pixels, multimodal spatial tokens interact with one another to construct the model’s understanding of the 3D scene. As a result, traditional 2D pruning strategies are not entirely suitable for this context. To address this challenge, we introduce AdaToken-3D, an adaptive strategy for pruning spatial tokens based on their spatial information contribution in 3D Large Multimodal Models (LMMs), achieving parameter reductions that far exceed the order of magnitude of conventional pruning methods. AdaToken-3D divides information contribution into intra-modal and inter-modal components to assess the criticality of spatial tokens at each stage. Our approach then removes a portion of the spatial tokens at the end of each stage according to a ratio determined by a contribution function based on the attention mechanism. This design leverages intra-modal contribution to preserve 3D positional information while progressively pruning the model to minimize redundancy in 3D LMMs.

To investigate spatial information flow, we extended prior analyses of attention inefficiency. As depicted in Fig\textcolor{red}{.2(A)}, 3D LMMs exhibit more severe visual redundancy than
their 2D counterparts. Through rigorous experimentation, we validate AdaToken-3D's dual efficacy: The LLaVA-3D-7B model with AdaToken-3D achieves a 60\% reduction of FLOPs while maintaining performance on core 3D benchmarks (SQA3D, ScanQA\cite{19}, ScanRefer\cite{22}, etc.), simultaneously serving as a plug-and-play module for
training acceleration that reduces GPU hours by 37\%. 
This framework establishes a methodological paradigm for cross-modal information analysis and tasks. In summary, our contributions are as follows:
\begin{itemize}
    \item We proposed AdaToken-3D, an adaptive spatial token optimization framework that dynamically prunes redundant tokens layer by layer according to its information contribution.
    \item We explored the information flow distribution in 3D LMMs by inefficient attention analysis, provided a new analysis paradigm for multimodal information of 3D scenes. 
    \item Our extensive experiments on LLaVA-3D demonstrated the computational efficiency and the high performance of our method. 
\end{itemize}

\section{RELATED WORK}

\subsection{\textbf{3D Large-Multimodal-Models}}

In the realm of 3D scene-level understanding, significant progress has been 
spurred by the development of Large-Vision-Language Models (LVLMs). Leveraging large-scale integrated 
multimodal data as spatial tokens for reasoning and 
verification in 3D Large-Multimodal-Models (3D 
LMMs) has become a prevalent research direction. 
GPT4Scene\cite{8} represents a notable approach. It extracts multi-view spatial features from video information and 
reconstructs the BEV map. This method effectively 
captures the spatio-temporal information in videos, 
providing a comprehensive view of the 3D scene. In 
contrast, LL3DA\cite{9} directly employs a scene-level 3D 
point cloud encoder to extract 3D scene representations. 
PARIS3D\cite{10} is another remarkable example. which 
specializes in the reasoning-based 3D part segmentation 
task. Given complex and implicit text queries, PARIS3D 
can accurately output the segmentation masks of 3D 
objects. Distinct from these methods, LLaVA-3D\cite{16} 
directly uses 3D position embeddings in combination with multi-view images as input. This unique 3D representation allows LLaVA-3D to rapidly adapt to 3D scene understanding. Moreover, it maintains a relatively 
simple architecture while still enabling effective multimodal input processing. This simplicity and efficiency 
make LLaVA-3D an attractive option for our research.

\begin{figure}[htpb]
    \centering
    \includegraphics[width=1\columnwidth]{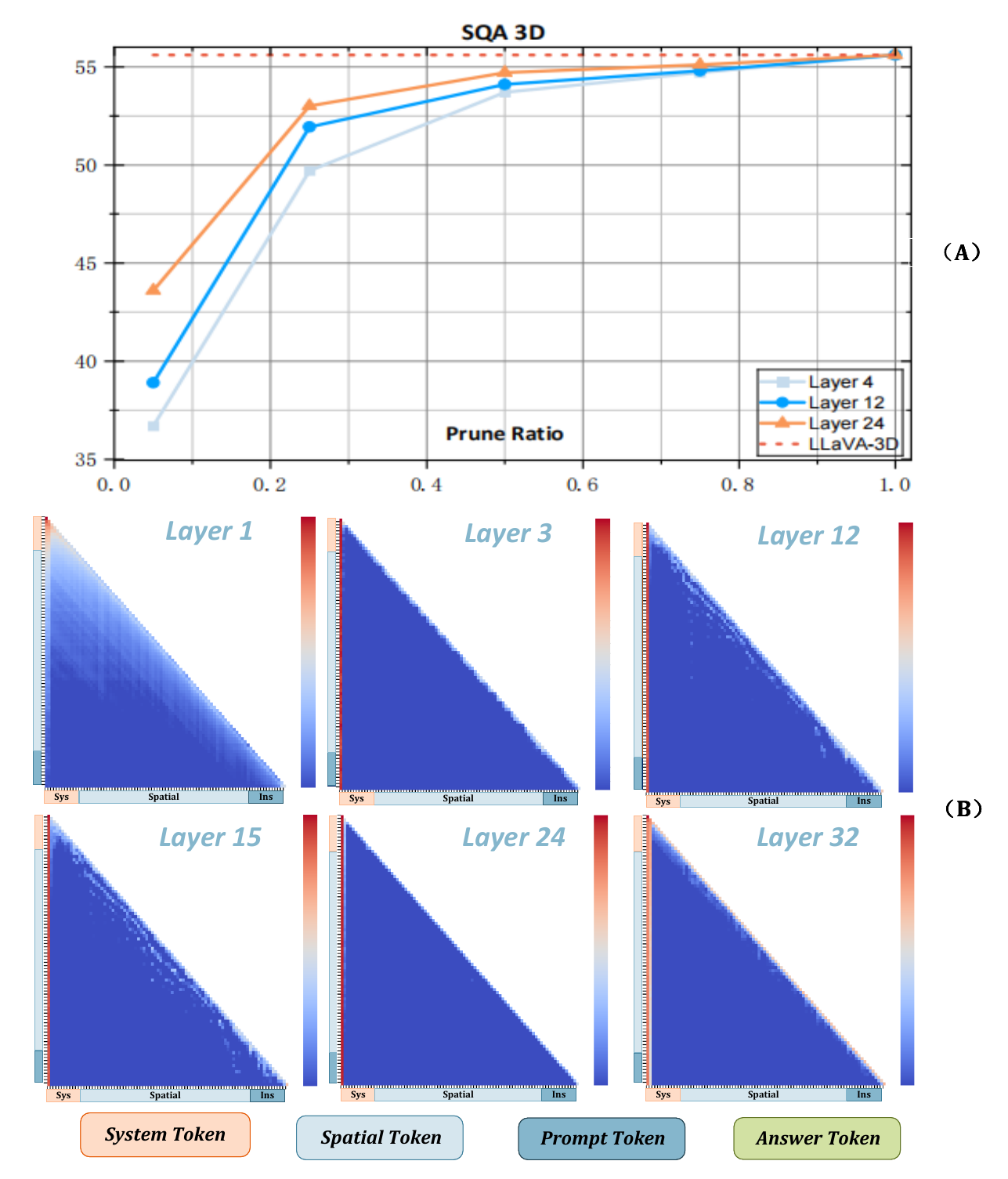}
    \caption{ Observations about Spatial Information Flow in 3D LMMs. (A) is the SQA3D performance of LLaVA-3D-7B with varying ration of pruned spatial tokens at different layers. Spatial-token ranking is based on its Attention Score. (B) is the attention map of system tokens,spatial tokens, prompt tokens and answer tokens. Spatial tokens, just like in 2D LLMs, get low attention allocation in deep layers(3-29). Notably, spatial tokens produce a large proportion of information redundancy in shallow layers, it confirms that the limited attention of the model cannot be spread across multiple input tokens. }
    \label{fig:st_collision_type}
\end{figure}
\subsection{\textbf{Information Redundancy in LVLMs}}
Previous studies have systematically addressed redundancy issues in LVLMs through two primary approaches: information flow optimization and computational efficiency enhancement. For architectural refinement, EAH\cite{11} introduces attention gate mechanisms to amplify shallow-layer image token propagation, effectively mitigating hallucinations without extensive retraining. Concurrently, PyramidDrop\cite{12} pioneers hierarchical token elimination, achieving 40\% training acceleration and 55\% FLOPs reduction for LLaVA-Next\cite{13}. Recent advances extend to dynamic computation paradigms, as exemplified by DynamicViT\cite{14} implementing entropy-based token prioritization and Ada-ViT\cite{15} adapting processing pathways through reinforcement learning. In contrast, AdaToken-3D makes a more in-depth study of redundancy in 3D-LLMs and proposes a dynamic spatial token reduction solution for the inference of LLMs.

\section{Methods}

\subsection{\textbf{Inefficient Spatial Attention in 3D LMMs}} 
As multimodal information injected into LMMs, it should be noted that higher performance will cause greater redundancy. In this section, we explored the necessity of spatial tokens in 3D LMMs from the analysis of the self-attention module and verification of the Pruning benchmark performance experiment.  

To obtain general rules, we choose LLava-3D-7B as the base model, which is known for its reasoning speed and high efficiency. LLava-3D-7B is 13 times faster than LEO in reasoning on multiple tasks. Based on multi-view 2D patch features 
$X_{2\text{D}}\in \mathbb{R}^{V\times c \times w \times h}$, LLava-3D combines it with  3D position embeddings $Pos^{\prime} \in \mathbb{R}^{V \times w \times h \times d}$ encoded by a 2-layer MLP, in order to get effective \textbf{spatial token representations} $X_{3 D} \in \mathbb{R}^{V \times w \times h \times d}$. Though its effective parameters have been greatly reduced compared to existing 3D LMMs, the tokens processed in LLava-3D are 1.1 times more than the LLaVA\cite{17}.

Considering LLava-3D consists of 32 layers, we utilize the attention map method brought by FastV to quantitatively study the redundancy of 3D models. The attention map records the attention score and importance of each token during the decoding process of LLava-3D.As is shown in Fig\textcolor{red}{.2(B)}, the information redundancy is still prominent in the deep layers of the model; spatial tokens which count for more than 80\% receive attention scores lower than 20\%. In the shallow layers, the lower part of spatial tokens receives hardly any attention; it verifies a model processing multiple tokens, its distribution of attention will be limited to a small amount of spatial tokens.On the contrast, most image tokens have high importance in the shallow layers in 2D LLMs. 

SQA3D \cite{c18} contains numerous 3DQA tasks requiring complex spatial reasoning that demands 3D LMMs to maintain fundamental generation capabilities while achieving comprehensive 3D scene understanding. The study systematically prunes spatial tokens at Layers 4,12 and 24 with varying ratios to investigate redundancy patterns across different reasoning stages. Building upon PyramidDrop\cite{12}, our pruning strategy employs lightweight attention computation to precisely identify instruction-critical spatial tokens. As illustrated in Fig\textcolor{red}{.2(B)}, the impact of spatial token removal on SQA3D performance diminishes with increasing network depth. This demonstrates that spatial tokens in shallow layers predominantly drive the model's reasoning process, while deeper layers exhibit higher redundancy.
\begin{figure*}[htpb]
    \centering
    \includegraphics[width=1.9\columnwidth]{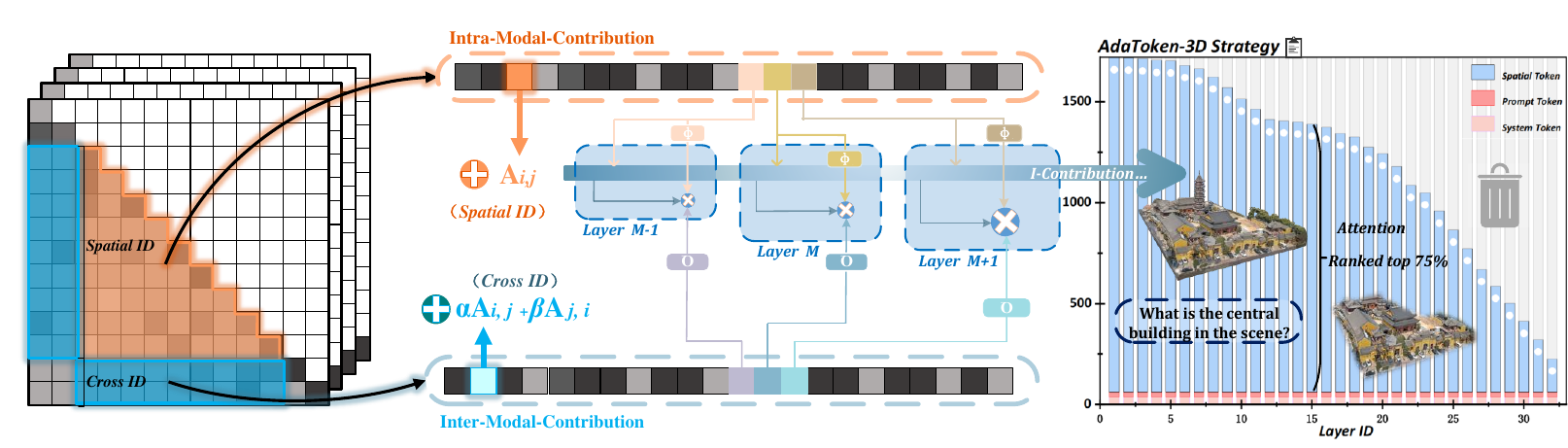}
    \caption{AdaToken-3D Pruning Strategy. We extract information in various forms using modal characteristics, and layer by layer, calculate the contribution of spatial tokens. This strategy, grounded in a lightweight pruning ratio mechanism, effectively reduces spatial tokens and selects crucial ones relevant to instructions.The pruning results on the right reflect the significance and smoothness of the strategy effect.}
    \label{fig:st_collision_type}
\end{figure*}
\subsection{\textbf{Calculation of Information Contribution}}
Previous research on information flow defines ‘information flow’ as the degree of influence of image, user, and system tokens on answer tokens\cite{1}, which has turned out to be an intuitive and effective quantitative analysis method. The key pixels in the images and the keywords in the prompts play a decisive role in the generating answer of the 2D LLMs. However, spatial tokens did not flow intuitively to answer tokens, for the position information contained in RGB-D interacts with each other to reconstruct the base scene, and original image information needs to be combined with the positions to render the scene back into the 3D worlds. To this end, we introduced Intra-Modal value $S_I^{\text {self }, i}$ and Inter-Modal value $S_I^{\text {cross }, i}$ to dynamically evaluate the contribution of each modal in layer \textit{i}.

Our method is inspired by FastV\cite{2}. The answer generation is sentence decoded from the answer tokens, thus we collect each token’s attention score distribution $\lambda$
 in different layers and weighted sum of attention scores of different types of tokens to calculate the information contribution of each layer. To better preserve the interactions and fusion between spatial tokens,
we compute the total attention allocation $S_{3D}^{\text {self }, i}$ to denote the total attention score spatial tokens received in one layer:
\begin{equation}
S_{3D}^{self,i}=\sum_{i=1}^{n}\lambda_{\mathrm{spa}}^{i}.
\label{eq:spa_energy}  
\end{equation}
To rigorously characterize the hierarchical propagation of spatial features in 3D visual understanding, we propose the Spatial Flow Value $F_{3D}^{i}$,a novel metric quantifying the dynamic information flux across successive decoding layers. Formally, for layer i, this spatio-temporal propagation mechanism is mathematically expressed as:
\begin{equation}
F_{\mathrm{3D}}^{i}=\sigma \cdot S_{\mathrm{3D}}^{self,i}+\gamma \cdot F_{\mathrm{3D}}^{i-1},
\label{eq:spa_energy}  
\end{equation}
 where $\sigma\in[0,1]$ represents the spatial attenuation factor controlling the current layer's contribution.To leverage past trends in spatial attention,$\gamma\in\mathbb{R}$ acts as the temporal persistence coefficient governing historical information retention.Next, we define Inter-modal value to be the weighted sum of the spatial-prompt and system-spatial interaction attention scores.It is used to indicate the contribution of the spatial modality to other modal information within the 3D LMMs.
 \begin{equation}
S_{3D}^{cross,i}=a_1 \cdot\sum_{i=1}^{n}\lambda_{s-p}^{i}+a_2 \cdot\sum_{i=1}^{n}\lambda_{s-s}^{i}.
\label{eq:spa_energy}  
\end{equation}
 
In our research, we calculate the information contribution $INF$ which extracts spatial modal interactive attention distribution characteristics and inter-layer historical change characteristics, exploring the importance of spatial tokens from multiple dimensions and emphasizing their role in 3D feature reconstruction.

\begin{equation}
{ 
 {INF}(i) = \exp\left(\frac{S_{3D}^{cross,i}}{\epsilon}\right) + \alpha_i \cdot F_{3D}^i+\log\left(1 + S_{3D}^{self,i}\right).
 }
\label{eq:enhanced}
\end{equation}

\subsection{\textbf{Adaptive Pruning Strategy for Spatial Tokens.}}
Contemporary pruning methodologies in multimodal large language models (MLLMs) predominantly rely on two major premises: (1) static attention scores sufficiently represent token importance, and (2) the pruning layers and ratios are predefined.Referring to our pruning experiment on 3D LMMs, these assumptions critically fail in 3D LMMs where cross-modal interactions exhibit  dependencies across layers.As is shown in Tab.1, our experiments reveal that naively pruning 40\% of tokens based on attention scores disrupts 42.3\%  performance on 3D VG tasks, demonstrating the fragility of such heuristic approaches in geometrically grounded reasoning scenarios.Thus,uniformly dropping spatial tokens across layers may lead to the loss of valuable
information while retaining unnecessary redundancy in the deeper layers.
\begin{table}[t]
\centering
\footnotesize
\scalebox{0.9}{
\begin{tabular}{@{}c@{\hspace{8pt}}cccc@{}} 
\toprule
 \multirow{2}{*}{Inference Strategy}& \multicolumn{2}{c}{\textbf{ScanRefer}}                                   & \multicolumn{2}{c}{\textbf{Scan2Cap}(Val)}\\ 
 & \multicolumn{1}{c}{\textit{Acc@0.25}}$\uparrow$ & \multicolumn{1}{c}{\textit{Acc@0.5}}$\uparrow$  & \multicolumn{1}{c}{\textit{R@0.5}}$\uparrow$  & \multicolumn{1}{c}{\textit{C@0.5}}$\uparrow$        \\ 
 \midrule
 LLaVA-3D\cite{16}& \textbf{54.1} & \textbf{42.2} & \textbf{63.41} & \textbf{79.21} \\ 
 FastV\cite{2}&   41.2  & 23.4 & 55.23 & 71.09 \\ 
 PyramidDrop\cite{12}& 43.9 & 32.7& 55.12 & 76.65 \\ 
 AdaToken-3D& \underline{53.2} & \underline{40.7} & \underline{60.23} & \underline{79.13} \\ 
\bottomrule
\end{tabular}
}
\caption{Evaluation results on the Benchmark of Scanrefer and Scan2Cap. Under the traditional 2D drop tokens strategy, the 3DVG capability of 3D LMMs is reduced by 26\% on average.}
\label{mme}
\end{table}
To address the problem of protecting multimodal information during the inference stage, we introduced AdaToken-3D. Based on the information contribution of each layer $INF(i)$,AdaToken-3D adjusts the retention ratio of each layer by introducing derivative constraints and dynamic optimization.A novel framework that dynamically adjusts layer-wise token retention through derivative-constrained optimization. The core innovation lies in its three-dimensional adaptive mechanism:

\begin{equation}
{
 {O^{pre}(i)} = \alpha\cdot\exp(-\beta(i-L))+M,
 }
\label{eq:enhanced}
\end{equation}
the proposed exponential decay mechanism incorporates three trainable parameters to adaptively regulate layer-wise information flow where initial attenuation amplitude $\alpha$, empirically constrained to $\alpha\in[0.5, 1.2]$, governs the magnitude disparity between the initial layer and subsequent decay regions, with elevated $\alpha$-values enhancing early-layer feature preservation. Attenuation rate $\beta$ , learned via back-propagation with $L_2$-regularization, determines the functional decay steepness  $\beta>1$ induces abrupt information reduction while  $\beta < 0.5$ facilitates gradual transitional phases. Attenuation center offset L modulates the phase shift of decay initiation, strategically positioned to align with empirical observations of critical layer distributions in transformer architectures, where intermediate layers typically exhibit peak information transfer capacities.
 
\begin{equation}
{
 {I_{norm}} = f({INF}),
 }
\label{eq:enhanced}
\end{equation}
\begin{equation} 
 \mathbf{L}=\sum_{i=1}^{n}({O^{pre}-I_{norm})^2}+\lambda \cdot\sum_{i=1}^{n}(\mathrm{\frac{\textit{d}\textit{O}}{\textit{d}\textit{i}}-\frac{\textit{dI}}{\textit{d}\textit{i}}})^2 .
 \label{eq:spa_energy}  
 \end{equation}
To optimize the model parameters, we employ a dual-constraint loss function $\mathbf{L}$ and we define the maximum and minimum normalization method as the function $f$, which simultaneously balances the accuracy of token ratio and the smoothness of inter-layer information flow. The optimization problem is formulated as a constrained nonlinear programming task, where boundary constraints are introduced to enforce a target global token reduction ratio. SLSQP iteratively approximates the Hessian matrix and enforces constraints through the Karush-Kuhn-Tucker (KKT) conditions, ensuring both feasibility and optimality of the solution. The resulting parameter is optimized to minimize $\mathbf{L}$ while maintaining the desired token reduction ratio 
 $O^{ratio,i}$ across all layers.

\textbf{AdaToken-3D ranking of spatial tokens}. The reasoning mechanism of the model needs to be thoroughly explored to identify the most suitable representation for the importance of spatial tokens. 3D Large Language Models (3D LMMs) are a type of multimodal language model that comprises a tokenizer and an $L$-layer transformer decoder $T$. The input to the transformer $T$ includes both spatial tokens and text tokens $s_0$. During the forward pass of these tokens $t_0$, we are able to derive the hidden states of the spatial tokens and text tokens at layer $L$. For instance: 
$$
s_L,t_L =T_L(s_{L-1},t_{L-1}).
$$

In order to boost the calculation efficiency and streamline the information flow, we focus on calculating the attention solely between all the spatial tokens and the final token of the instruction (hereafter denoted as $t^{end}$, which represents the last instruction token).

Subsequently, we obtain the key states of the spatial tokens, denoted as $K^s=K_L(s_L)$, and the query state of the last prompt token, denoted as $Q^t=Q_L(t_L)$. Here, $K_L$ and  $Q_L$ are the key matrix and the query matrix, respectively, derived from the self-attention block of $T$. We compute the similarity $V_j$ and consider it as the ranking indicator for the spatial tokens, as follows:
$$V_j=Q^t \times (K^s)^T.
$$

Following our observation in Sec 3.2, our pruning strategy calculates the theoretical retention ratio $O^{ratio,i}$ of each layer based on the information contribution.After that,we reduced the corresponding number of tokens which ranked last as its $V_j^n$. Through the strategy of AdaToken-3D, the spatial token number decreases exponentially layer by layer, and the tokens will be dropped more quickly in the deeper layer.Compared with the fixed ratio strategy, AdaToken-3D will maximize the retention of 3D reasoning capabilities and reserve the integrity of the reasoning process.

\textbf{AdaToken-3D Computation Cost Estimation}, Here, we analyze the computational cost of AdaToken-3D from two aspects: one is the computational overhead introduced by AdaToken-3D, and the other is the computational cost of spatial tokens saved by AdaToken-3D. The additional computational cost introduced by AdaToken-3D mainly lies in the similarity calculation of spatial token sorting and the calculation of information contribution of each layer. Due to our design, the calculation is only performed between a query token and the spatial tokens, so the overhead caused by its computational complexity during the forward propagation process is relatively small. The computational cost saved by AdaToken-3D, spatial tokens can be expressed as a linear function with a constant factor. Therefore, when using AdaToken-3D with n layers and a certain proportional relationship,the the token utilization rate increased by 23\%, the FLOPs during the inference mode is reduced by 63\%.

\section{Experiments}
In this section, we conduct extensive evaluations to examine the effectiveness and generalization of AdaToken-3D. To begin with, we introduce the experiment implementation details (Sec.\textcolor{red}{4.1}). Then, we compare our method’s 3D scene understanding capability (Sec.\textcolor{red}{4.2}) and evaluate its efficiency in 3D tasks (Sec.\textcolor{red}{4.3}). Finally, we demonstrate that AdaToken-3D could preserve the 2D image understanding capabilities of LLaVA-3D and the method is still applicable to 2D LLMs(Sec.\textcolor{red}{4.4}).
\begin{table*}[t]
\centering
\footnotesize
\scalebox{1}{
\begin{tabular}{@{}c@{\hspace{8pt}}ll|llccc|cc@{}} 
\toprule
 & &&  \multicolumn{5}{c}{\textbf{ScanQA}}&\textbf{SQA3D}\\ 
  \textit{Methods}&\textit{Infer-mode}&\textit{Ratio}&  \textit{C}$\uparrow$& \textit{B4}$\uparrow$ &\multicolumn{1}{c}{\textit{M}}$\uparrow$ & \multicolumn{1}{c}{\textit{R}}$\uparrow$ & \multicolumn{1}{c}{\textit{EM@1}}$\uparrow$ & \multicolumn{1}{c}{\textit{EM@1}}$\uparrow$       \\ 
\midrule
 \textit{LEO\cite{31}}&\textit{Vanilla} &\textit{100\%}&  \textbf{101.4}& 13.2&\underline{20.0}& \underline{49.2}& 24.5& 50.0\\ 
 \textit{Scene-LLM\cite{3}}&\textit{Vanilla} &\textit{100\%}&    80& 12.0&16.6& 40.0& 27.2& 54.2\\ 
 \textit{LLaVA-3D\cite{16}}&\textit{Vanilla} &\textit{100\%}&  91.7& \textbf{14.5}&\textbf{20.7}& \textbf{50.1}& 27.0& \textbf{55.6}\\ 
 \midrule
 \textit{LLaVA-3D}&\textit{PDrop\cite{12}} &\textit{10\%}&  80& 8.2&15.6& 39.9& 28.6 & 52.7\\
 \textit{--}& \textit{PDrop}&\textit{20\%}& 87.6& 10.3& 16.9& 43.3& 28.6&53.6\\
 \textit{--}& \textit{PDrop}&\textit{40\%}& 91.0& 11.3& 17.0& 43.8& 28.2&54.7\\
 \midrule
 \textit{LLaVA-3D}&  \textit{Ada-3D}&\textit{10\%}& 90.7& 11.5& 17.2& 43.7& 28.5&53.5\\
 \textit{--}&  \textit{Ada-3D}&\textit{20\%}& 90.6& 12.1& 16.8& 44.7& 29.8&54.7\\
 \textit{--}&  \textit{Ada-3D}&\textit{40\%}& \underline{91.8}& \underline{13.3}& 17.6& 44.7&\textbf{ 30.0}&\underline{54.8}\\ 
\bottomrule
\end{tabular}
}
\caption{\textbf{Inference Acceleration Performance on 3D QA Tasks}
 “C” represents “CIDEr”, “B-4” represents “BLEU-4”, “M” represents “METEOR”, “R” represents “ROUGE”, “Sim” represents sentence similarity, and “EM@1” represents top-1 exact match. AdaToken-3D prunes most of the spatial tokens without sacrificing performance in 3D QA tasks. Notably, compared with traditional 2D pruning methods like PyramidDrop, the performance of AdaToken-3D is less affected by the pruning ratio.}
\label{mme}
\end{table*}

\subsection{\textbf{Setup}}
\textbf{Experiment models} We conduct comprehensive experiments to validate the effectiveness of our proposed AdaToken-3D framework across various large language models (LLMs) and input resolutions. Specifically, we evaluate LLaVA-3D-7B on 3D scene understanding tasks while comparing its 2D performance against established baselines including LLaVA-NeXT-Vicuna-7B \cite{13} and LLaVA-1.5-Vicuna-7B\cite{28}. Its framework builds upon LLaVA, the predominant open-source LLM backbone in the research community, which employs a streamlined architecture that effectively aligns visual features from CLIP encoders with language model inputs through a feature projection module. The key character of LLaVA-3D lies in its ability to extend 2D visual-language understanding to 3D spatial reasoning without performance degradation in conventional 2D tasks. Our experimental results demonstrate that AdaToken-3D successfully preserves original 3D comprehension capabilities while generalizing to 2D LLMs.

\textbf{Implementation Details} Leveraging the token drop ratio and the information contribution $INF(i)$ of each layer, we utilize Adatoken-3D to determine the pruning ratio for each layer, progressively reducing the count of spatial tokens layer by layer. All experiments are executed on a setup comprising 8 × 40G A6000 GPUs. During the accelerated inference time evaluation, we have the flexibility to adjust the ratio $\alpha$ to regulate the extent to which spatial tokens are pruned, with a default setting of $\alpha$=0.8. Our 3D evaluation benchmark primarily relies on the Scannet Datasets\cite{29}, where we have selected 32 scenarios and rigorously tested thousands of QA pairs within each benchmark. We plan to disclose the specific scenarios selected later to ensure the transparency and objectivity of our experimental findings.

\textbf{Benchmarks} To comprehensively assess 3D scene understanding capabilities, we employ benchmarks spanning multiple task categories. The SQA3D benchmark expands into dynamic 3D scenes through the use of 19K GPT-4-generated questions, alongside the Scannet-SQA dataset, which boasts 33.4K human-annotated questions. ScanQA serves as a benchmark offering 27K question-answer pairs to evaluate spatial relationships and navigation planning skills. As depicted in Tab\textcolor{red}{.1}, the ScanRefer benchmark is employed to report on the performance of localizing target objects within 3D scenes using natural language descriptions. Scan2Cap\cite{23} challenges models to describe an object's appearance and its spatial relations with neighboring objects, outputting the corresponding 3D bounding box. MM-Vet\cite{24} harnesses GPT-4\cite{30} for a six-dimensional evaluation of LMM capabilities. Additionally, traditional 2D capability benchmarks such as MME\cite{25} and MMB\cite{26} are also leveraged. Lastly, VQA-v2\cite{27} is included in our evaluation suite.
\begin{table}[t]
\centering
\footnotesize
\scalebox{0.9}{
\begin{tabular}{@{}c@{\hspace{8pt}}|cccc@{}} 
\toprule
 \multirow{2}{*}{Model} & \multicolumn{2}{c}{\textbf{Inference(Scene0011-00)}}& \multicolumn{2}{c}{\textbf{SQA3D(1000QA-pairs)}}\\ 
  & \multicolumn{1}{c}{\textit{Total-time}(s)} & \multicolumn{1}{c}{\textit{FLOPs}(T)} & \multicolumn{1}{c}{\textit{Total-time}(s)} & \multicolumn{1}{c}{\textit{Latency}}       \\ 
 \midrule
 \textit{LLaVA-3D\cite{16}}& 5.957& 11.46& 550& 72\%\\ 
 \textit{PyramidDrop\cite{12}}& 5.026& 5.23& 372& 16.6\%\\ 
 \textit{AdaToken-3D}& \textbf{4.439}& \textbf{4.57}& \textbf{319}& -\\ 
\bottomrule
\end{tabular}
}
\caption{The inference efficiency comparison between pruning strategy and vanilla decoding. We evaluate the inference efficiency of single complex inference and multiple simple inference. }
\label{mme}
\end{table}
\subsection{\textbf{Performance of AdaToken-3D in inference}}
3DQA tasks demand that a model generate responses to natural-language queries about spatial aspects. To assess whether 3D large-Multimodal-Models (LMMs) can maintain an understanding of 3D scenes, we employ a diverse range of benchmarks for comparative verification.

As shown in Tab\textcolor{red}{.2}, we set three groups of discarded token ratios. At each drop ratio, AdaToken-3D performs on a par with LLaVA-3D. This not only reveals the substantial redundancy within 3D LMMs but also, indirectly, validates the consistency of our spatial token ranking strategy.
When compared to the 2D pruning strategy PyramidDrop, at a ratio of 40\%, their performances are similar. However, as the ratio decreases, the performance of PyramidDrop declines significantly. In contrast, AdaToken-3D remains relatively stable and has not yet reached its minimum effective ratio.

In summary, traditional 2D pruning strategies limit the potential for reducing redundancy. This is because they rely on predefined drop ratios and overlook the information flow in multimodal data. In contrast, our AdaToken-3D can prune up to 90\% of the spatial tokens without degrading the performance of LLaVA-3D.It should be noted that the process of reducing redundancy is also the process of reducing 3D LMMs' illusion; less spatial tokens also mean less useless tokens.

\textbf{The AdaToken-3D strategy also can be used as a training strategy.} Adatoken-3D is proposed to reduce the redundancy within spatial tokens, and as we observed above, it enjoys higher speedup with the increase of the drop token ratio.
With the increased drop ratio, Adatoken-3D reaches a higher speedup that only 73 GPU hours are used for training. which could reduce almost 45\% training time of the vanilla LLaVA-3D-7B. 
\subsection {\textbf{Assessing the Inference Efficiency of AdaToken-3D}}To rigorously quantify the computational advantages and real-time performance capabilities of AdaToken-3D, we maintain the identical experimental configurations outlined in Section\textcolor{red}{4.2}. In this context, we meticulously evaluate the efficiency of our proposed method. As illustrated in Tab\textcolor{red}{.3}, we seamlessly integrate the dynamic pruning strategy into the inference phase of the baseline model, juxtaposing its performance against the inference acceleration technique known as PyramidDrop.

Our findings on the LLaVA-3D benchmark underscore the remarkable superiority of AdaToken-3D over PDrop in terms of inference speed. Precisely, we achieve a stunning 72\% reduction in latency across multiple straightforward inference tasks, surpassing PDrop by a notable margin of 17.6\%. Furthermore, our approach demonstrates a substantial decrease in FLOPs utilization during complex reasoning tasks, amounting to a 60\% reduction.

When benchmarked against the baseline model, AdaToken-3D consistently maintains comparable performance levels across a wide array of benchmarks, solidifying its efficacy and versatility. These results not only highlight the significant computational gains but also underscore the practicality of our method in real-world applications requiring swift and accurate inference.
\subsection {\textbf{Comparison on 2D benchmarks}}

As demonstrated in Tab\textcolor{red}{.4}, we have directly implemented the AdaToken-3D strategy within the inference phase of the LLaVA-3D model when tackling 2D tasks. This implementation underscores our method's adaptability and versatility, extending its applicability beyond 3D data to include two-dimensional contexts.

In comparing AdaToken-3D with conventional 2D pruning strategies, it is noteworthy that, despite achieving a comparable token reduction ratio, our method exhibits performance levels on par with PyramidDrop and FastV, underscoring the efficiency and efficacy of our proposed strategy.Moreover, when evaluating AdaToken-3D on the same backbone model across diverse benchmarks, we observe a notable superiority in both Visual Question Answering (VQA) and MMB. By dynamically selecting and processing only the most important tokens, AdaToken-3D offers a promising solution to the challenge of balancing performance and efficiency in visual reasoning tasks.
\begin{table}[t]
\centering
\footnotesize
\scalebox{0.9}{
\begin{tabular}{@{}c@{\hspace{8pt}}l|cccc@{}} 
\toprule
 \multirow{2}{*}{Model} &  \multirow{2}{*}{Infer}&\multirow{2}{*}{VQA$\uparrow$}&\multirow{2}{*}{MMB$\uparrow$}& \multirow{2}{*}{MME$\uparrow$}&\multirow{2}{*}{MM-Vet$\uparrow$}\\ 
  &  &\multicolumn{1}{c}{} & \multicolumn{1}{c}{} & \multicolumn{1}{c}{} & \multicolumn{1}{c}{}       \\ 
 \midrule
 \multirow{2}{*}{LLaVA-1.5-7B\cite{17}}&  \textit{FastV}&57.0& 56.3& 1475.3& 30.8\\
 &  \textit{PDrop}&57.4& 58.9& 1498.3&\textbf{31.3}\\ 
 \midrule
 \multirow{2}{*}{LLaVA-NeXT-7B\cite{13}}&  \textit{FastV}&\underline{66.0}& 59.8& 1483.1& -\\
 &  \textit{PDrop}&\textbf{67.1}& 60.0& \textbf{1532.9}&-\\
 \midrule
 \multirow{2}{*}{LLaVA-3D-7B\cite{16}}&  \textit{Ada-3D}&58.1& \textbf{65.0}& \underline{1501.8}&30.8\\ 
 &  \textit{PDrop}&57.9& \underline{64.9}& 1501.5& \underline{30.9}\\ 
\bottomrule
\end{tabular}
}
\caption{\textbf{Comparisons with pruning strategy on zero-shot 2D benchmarks.}Adatoken 3D still has a relatively impressive performance in 2D tasks and has good versatility.}
\label{mme}
\end{table}
\section{CONCLUSIONS}
This work addresses critical inefficiencies in 3D large multimodal models (LMMs) by proposing AdaToken-3D, an adaptive spatial token pruning framework that optimizes computational redundancy while preserving 3D scene understanding capabilities. By analyzing spatial information flow and introducing intra- and inter-modal contribution metrics. The study reveals distinct spatial information dynamics in 3D LMMs compared to 2D counterparts, offering new insights into cross-modal reasoning mechanisms. These advancements establish a foundational paradigm for developing efficient, human-aligned spatial intelligence systems and accelerating progress toward embodied AI.

\addtolength{\textheight}{-12cm}   






\begin{thebibliography}{99}

\bibitem{1} Zhang, Xiaofeng et al. “From Redundancy to Relevance: Information Flow in LVLMs Across Reasoning Tasks.” (2024).
\bibitem{2} Chen, Liang et al. “An Image is Worth 1/2 Tokens After Layer 2: Plug-and-Play Inference Acceleration for Large Vision-Language Models.” European Conference on Computer Vision (2024).
\bibitem{3} Fu, Rao et al. “Scene-LLM: Extending Language Model for 3D Visual Understanding and Reasoning.” ArXiv abs/2403.11401 (2024): n. pag.
\bibitem{4} Chu, Xiangxiang et al. “MobileVLM : A Fast, Strong and Open Vision Language Assistant for Mobile Devices.” ArXiv abs/2312.16886 (2023): n. pag.
\bibitem{5} Huang, Haifeng et al. “Chat-Scene: Bridging 3D Scene and Large Language Models with Object Identifiers.” Neural Information Processing Systems (2023).
\bibitem{6} Huang, Jiangyong et al. “An Embodied Generalist Agent in 3D World.” ArXiv abs/2311.12871 (2023): n. pag.
\bibitem{7} Lu, Haoyu et al. “DeepSeek-VL: Towards Real-World Vision-Language Understanding.” ArXiv abs/2403.05525 (2024): n. pag.
\bibitem{8} Qi, Zhangyang et al. “GPT4Scene: Understand 3D Scenes from Videos with Vision-Language Models.” ArXiv abs/2501.01428 (2025): n. pag.
\bibitem{9} Chen, Sijin et al. “LL3DA: Visual Interactive Instruction Tuning for Omni-3D Understanding, Reasoning, and Planning.” 2024 IEEE/CVF Conference on Computer Vision and Pattern Recognition (CVPR) (2023): 26418-26428.
\bibitem{10} Kareem, Amrin et al. “PARIS3D: Reasoning-based 3D Part Segmentation Using Large Multimodal Model.” ArXiv abs/2404.03836 (2024): n. pag.
\bibitem{11} Zhang, Xiaofeng et al. “Seeing Clearly by Layer Two: Enhancing Attention Heads to Alleviate Hallucination in LVLMs.” ArXiv abs/2411.09968 (2024): n. pag.
\bibitem{12} Xing, Long et al. “PyramidDrop: Accelerating Your Large Vision-Language Models via Pyramid Visual Redundancy Reduction.” ArXiv abs/2410.17247 (2024): n. pag.
\bibitem{13} Li, Feng et al. “LLaVA-NeXT-Interleave: Tackling Multi-image, Video, and 3D in Large Multimodal Models.” ArXiv abs/2407.07895 (2024): n. pag.
\bibitem{14} Rao, Yongming et al. “DynamicViT: Efficient Vision Transformers with Dynamic Token Sparsification.” ArXiv abs/2106.02034 (2021): n. pag.
\bibitem{15} Meng, Lingchen et al. “AdaViT: Adaptive Vision Transformers for Efficient Image Recognition.” 2022 IEEE/CVF Conference on Computer Vision and Pattern Recognition (CVPR) (2021): 12299-12308.
\bibitem{16} Zhu, Chenming et al. “LLaVA-3D: A Simple yet Effective Pathway to Empowering LMMs with 3D-awareness.” ArXiv abs/2409.18125 (2024): n. pag.
\bibitem{17} Liu, Haotian et al. “Visual Instruction Tuning.” ArXiv abs/2304.08485 (2023): n. pag.
\bibitem{18} Ma, Xiaojian et al. “SQA3D: Situated Question Answering in 3D Scenes.” ArXiv abs/2210.07474 (2022): n. pag.
\bibitem{19}Azuma, Daich et al. “ScanQA: 3D Question Answering for Spatial Scene Understanding.” 2022 IEEE/CVF Conference on Computer Vision and Pattern Recognition (CVPR) (2021): 19107-19117.
\bibitem{20} Chen, Zhe et al. “Expanding Performance Boundaries of Open-Source Multimodal Models with Model, Data, and Test-Time Scaling.” ArXiv abs/2412.05271 (2024): n. pag.
\bibitem{21}Bai, Jinze et al. “Qwen-VL: A Versatile Vision-Language Model for Understanding, Localization, Text Reading, and Beyond.” (2023).
\bibitem{22}Chen, Dave Zhenyu et al. “ScanRefer: 3D Object Localization in RGB-D Scans using Natural Language.” ArXiv abs/1912.08830 (2019): n. pag.
\bibitem{23}Chen, Dave Zhenyu et al. “Scan2Cap: Context-aware Dense Captioning in RGB-D Scans.” 2021 IEEE/CVF Conference on Computer Vision and Pattern Recognition (CVPR) (2020): 3192-3202.
\bibitem{24}Yu, Weihao et al. “MM-Vet: Evaluating Large Multimodal Models for Integrated Capabilities.” ArXiv abs/2308.02490 (2023): n. pag.
\bibitem{25}Fu, Chaoyou et al. “MME: A Comprehensive Evaluation Benchmark for Multimodal Large Language Models.” ArXiv abs/2306.13394 (2023): n. pag.
\bibitem{26}Liu, Yuanzhan et al. “MMBench: Is Your Multi-modal Model an All-around Player?” European Conference on Computer Vision (2023).
\bibitem{27}Goyal, Yash et al. “Making the V in VQA Matter: Elevating the Role of Image Understanding in Visual Question Answering.” International Journal of Computer Vision 127 (2016): 398 - 414.
\bibitem{28}Kassem, Aly M. et al. “Alpaca against Vicuna: Using LLMs to Uncover Memorization of LLMs.” ArXiv abs/2403.04801 (2024): n. pag.
\bibitem{29}Dai, Angela et al. “ScanNet: Richly-Annotated 3D Reconstructions of Indoor Scenes.” 2017 IEEE Conference on Computer Vision and Pattern Recognition (CVPR) (2017): 2432-2443.
\bibitem{30}Achiam, OpenAI Josh et al. “GPT-4 Technical Report.” (2023).
\bibitem{31}Huang, Jiangyong et al. “An Embodied Generalist Agent in 3D World.” ArXiv abs/2311.12871 (2023): n. pag.




\end{thebibliography}
\end{document}